\documentclass[onecolumn,showpacs,preprintnumbers, 
               superscriptaddress,
               eqsecnum,nofootinbib]{revtex4}
\usepackage{graphicx, fancybox}
\usepackage{amsmath,amssymb}
\usepackage[dvipdfm, colorlinks=true, pdfstartview=FitV, linkcolor=red, citecolor=blue, urlcolor=blue]{hyperref} 
\usepackage{color} 
\usepackage{bm}
\newcommand{\vect}[1]{\mathbf{#1}}

\newcommand{\sgn}{\mathrm{sgn}}

\newcommand{\vp}{\vect{p}}
\newcommand{\vq}{\vect{q}}
\newcommand{\vk}{\vect{k}}

\newcommand{\vx}{\vect{x}}
\newcommand{\vy}{\vect{y}}

\newcommand{\vj}{\vect{j}}

\newcommand{\vzero}{\vect{0}}

\newcommand{\muf}{\mu_f}
\newcommand{\mub}{\mu_b}
\newcommand{\Nf}{N_f}

\newcommand{\nf}{n_f}
\newcommand{\nb}{n_b}
\newcommand{\rhof}{\rho_f}
\newcommand{\rhob}{\rho_b}
\newcommand{\Ubb}{U_{bb}}
\newcommand{\Ubf}{U_{bf}}
\newcommand{\ef}{\epsilon^f_\vk}
\newcommand{\eb}{\epsilon^b_\vk}

\newcommand \beq{\begin{eqnarray}}
\newcommand \eeq{\end{eqnarray}} 
\newcommand{\nn}{\nonumber\\ }
\def\bra#1{\langle#1\vert}
\def\ket#1{\vert#1\rangle}
\newcommand{\rmd}{{\rm d}}
\newcommand{\rme}{{\rm e}}
\newcommand{\del}{\partial}

\begin{document}

\title{Spectral properties of the Goldstino  in supersymmetric Bose-Fermi mixtures
}

\author{Jean-Paul Blaizot}
\affiliation{Institut de Physique Th\'eorique, CEA/Saclay,  CNRS/UMR 3681,  F-91191 Gif-sur-Yvette Cedex, France}

\author{Yoshimasa Hidaka}
\affiliation{Theoretical Research Division, Nishina Center,
RIKEN, Wako 351-0198, Japan}

\author{Daisuke Satow}
\email{sato@fbk.eu}
\affiliation{European Center for Theoretical studies in Nuclear Physics and Related Areas (ECT*),
Strada delle Tabarelle 286, I-38123 Villazzano (TN) Italy}

\begin{abstract} 
We evaluate the spectral properties of the Goldstino in a Bose-Fermi mixture of cold atoms and molecules whose dynamics is governed by a supersymmetric hamiltonian. 
Model independent results are obtained from sum rules obeyed by the spectral function. We carry out specific calculations, at zero and finite temperature, using the Random Phase Approximation, and obtain in particular analytic expressions for the dispersion relation at small momentum. These explicit calculations allow us to pin down more precisely the features of the Goldstino that can be attributed to the supersymmetry alone,  together with its spontaneous breaking.  The anomalous large effect of the Fermi sea at moderate and large momenta is emphasized. 
\end{abstract} 

\date{\today}

\pacs{11.30.Pb, 
67.85.Pq	
}

\maketitle

\section{Introduction}
\label{sec:intro}

It has been suggested recently that some special Bose-Fermi mixtures of cold atoms and molecules on optical lattices could be prepared in such a way as they exhibit symmetry under the interchange of bosons and fermions (see \cite{Yu:2007xb,Yu:2010zv}, and references therein). 
In fact, the underlying algebraic structure of the hamiltonian that governs the dynamics of such systems presents some similarity with the supersymmetry of high energy physics~\cite{Wess:1974tw}.  
Although, in the present case, the supersymmetry does not involve space time coordinates, this analogy has triggered many theoretical works on the subject \cite{Yu:2007xb,Yu:2010zv,Shi:2009ak, Lai:2015fia, Bradlyn:2015kca}. 
In particular, it has been suggested that, since supersymmetry is broken at finite temperature and/or density, an analog of a Nambu-Goldstone excitation, dubbed the ``Goldstino''~\cite{Kratzert:2003cr},  should appear in the spectrum~\cite{Yu:2007xb}. 

The study of this Goldstino is interesting   for a variety of reasons.  Such modes have been studied in genuine supersymmetric relativistic quantum field theories or models~\cite{Kratzert:2003cr}, and it is  worthwhile to compare the properties of these modes with those that can be produced in cold atom systems. Collective phenomena where bosons are turned into fermions are also expected in the high temperature quark-gluon plasma, even though the supersymmetry there is only approximate\footnote{Other models with only approximate supersymmetry, in which a Goldstino excitation emerges, have been studied, see for instance~\cite{Satow:2013oya, Blaizot:2014hka}.}, and its role rather subtle~\cite{Hidaka:2011rz}. 
 The connection between the supersymmetry breaking  (spontaneous or explicit)~\cite{VanHove:1982cc},  and the existence of the Goldstino,  raises also interesting questions~\cite{Das:1978rx}. Finally, 
  the Goldstino is an example of a collective excitation carrying fermionic quantum number\footnote{Another such an excitation exists in the quark-gluon plasma, the so-called ``plasmino'', see \cite{Blaizot:2001nr,blaizot-mu}, and references therein. }, a rather unusual feature in many body systems where most of well known collective excitations have a bosonic character.  All these issues provide motivation for the present study, which is however more limited in scope.

Our discussion builds on the recent suggestion presented in Ref.~\cite{Shi:2009ak}, where an ingenious experimental setup was proposed to obtain evidence for the Goldstino. This setup involves (fermionic) atoms and (bosonic) molecules  placed in an optical lattice. The molecule is made of two fermions that exist in two species, with one species being essentially inert. The latter will then be ignored in our discussion, as well as many other details that are important for the experimental realization, but not for our present purpose. Our study will be based on a simple continuum hamiltonian inspired by the original setup of Ref.~\cite{Shi:2009ak}, and whose parameters are chosen so that it exhibits supersymmetry. Our goal is to study the spectral properties of the Goldstino, how these are affected by the temperature, and to which extent these are determined by symmetry consideration alone. We shall discover that the supersymmetry predicts indeed the existence of a sharp excitation at zero momentum at an energy given by  the difference of the fermion and boson chemical potentials, but that this mode is strongly modified by the presence of fermions at finite momentum.

This paper is organized as follows: In the next section, we  introduce a specific model of a Bose-Fermi mixture of cold atoms, that can exhibit a special (super)symmetry under the interchange of  bosons into fermions. We review some consequences of this supersymmetry in particular on the degeneracies of the excitation spectrum~\cite{Yu:2007xb}. 
Finite density breaks the supersymmetry, and the properties of the associated Nambu-Goldstone collective mode, the Goldstino, are discussed in a model independent way, in particular with the help of sum rules.  
Then in Sec.~\ref{sec:spectrum}, we present results of explicit calculations of the collective excitations using the Random Phase Approximation (RPA). 
The spectral function is analyzed in generic situations, at zero and finite temperature. 
We obtain analytic expressions for the dispersion relation and the strength of the Goldstino at small momentum.
The specific effect of the Fermi sea on the finite momentum excitations is emphasized. 
The last section summarizes our conclusions and point out possible extensions of the present work.

\section{A simple supersymmetric model and the Goldstino excitation}
\label{sec:Intro-2}

Although most of the arguments that we shall develop have a broader range of applicability, we shall base our discussion on the following specific hamiltonian $H=H_0+V$ in $d$ dimensions, with 
\begin{align} 
\label{eq:Hkin-continuum}
H_0&= -t_b\int d^d\vx\, b^\dagger (\vx)\nabla^2 \,b(\vx) -t_f\int d^d\vx\, f^\dagger (\vx) \nabla^2 f(\vx) \nn
V&= \int d^d\vx \biggl(\frac{U_{bb}}{2}b^\dagger(\vx) b^\dagger(\vx) b(\vx) b(\vx)
+U_{bf} b^\dagger(\vx) b(\vx) f^\dagger(\vx) f(\vx)\biggr) \nn
&= \int d^d\vx \biggl(\frac{U_{bb}}{2} n_b(\vx)(n_b(\vx)-1)+U_{bf}n_b(\vx)n_f(\vx)\biggr) ,
\end{align}
where $b(\vx)$ and $f(\vx)$ denote respectively the annihilation operators of bosons and fermions, with $b^\dagger(\vx)$ and $f^\dagger(\vx)$ the corresponding creation operators, and $n_b(\vx)$, $n_f(\vx)$ are, respectively, the local densities of bosons and fermions:
\beq
n_b(\vx)\equiv b^\dagger(\vx) b(\vx),\qquad n_f(\vx)\equiv  f^\dagger(\vx) f(\vx).
\eeq
We assume that the bosons have zero spin, and that only one spin component of the fermions is active, so that the spin degrees of freedom can be forgotten. 
This hamiltonian can be viewed as the continuum version of the  lattice model suggested in Ref.~\cite{Shi:2009ak}, with on-site interactions between the bosons and the fermions. This is the origin of the notation $t_b$, $t_f$ for the kinetic (hopping) terms of the boson and the fermion. There is no local fermion-fermion interaction,  because all fermions have the same spin component. 

\subsection{Supersymmetry and degeneracies}

The hamiltonian (\ref{eq:Hkin-continuum}) possesses  symmetries that we now examine. First, it is easily verified that it commutes with the particle number operators $N_b$ and $N_f$:
\beq
[H,N_b]=[H,N_f]=0,
\eeq
with
\beq
 \qquad N_b=\int d^d\vx\, n_b(\vx),\qquad 
 N_f=\int d^d\vx\,n_f(\vx), \\
 \nonumber
  n(\vx)\equiv  n_f(\vx) + n_b(\vx), \qquad
 N\equiv N_b+N_f,
\eeq
where we have introduced the total particle density $n(\vx)$ and number $N$ for future convenience.
Next, let us consider the  following operator
\beq
Q\equiv \int d^d \vx\, q(\vx),
\eeq
where $q(\vx)$ annihilates one boson and creates one fermion at point $\vx$ 
\begin{align}
q(\vx)\equiv f^\dagger(\vx) b(\vx) .
\end{align}
By using the canonical commutation and anticommutation relations
\beq\label{comrel}
[b(\vx),b^\dagger(\vy)]=\delta^{(d)}(\vx-\vy),\qquad 
\{f(\vx),f^\dagger(\vy)\}=\delta^{(d)}(\vx-\vy),
\eeq
one easily establishes the following algebra
\beq\label{algebra}
&&[Q,N_b]=Q,\qquad [Q,N_f]=-Q,\nn
&&[Q,N_b+N_f]=0,\qquad [Q,N_b-N_f]=2Q,\nn
&&\{Q,Q^\dagger\}=N.
\eeq
Note also the important relation 
\beq
Q^2=0.
\eeq
We shall refer to the operator $Q$ as the ``supercharge'', and to the algebra above as to a supersymmetric (SUSY) algebra. We emphasize however that this algebra differs from that considered in the context of high-energy physics ~\cite{Wess:1974tw}:
The latter is related to the spacetime symmetries, while the former is not. The present supersymmetry is best viewed as an internal  symmetry under the exchange of bosons into fermions and vice versa. When $t_b=t_f$ and  $\Ubb=\Ubf$, the hamiltonian $H=H_0+V$ commutes with the supercharge, and these conditions will be assumed to hold throughout this paper. We shall then set $t_{\rm h}=t_f=t_b$ and  $U=U_{bb}=U_{bf}$.

The commutation relations (\ref{algebra}) above imply degeneracies among the eigenstates of the Hamiltonian. We can label these eigenstates by the eigenvalues $N_b$ and $N_f$ of the number operators. Thus we denote a typical eigenstate by $\ket{n, N_b,N_f}$, with $n$ standing for all the other quantum numbers that are necessary to completely specify the state. The supercharge acting on such a state yields  
\beq\label{superNbNf}
&&Q\ket{n, N_b,N_f}=\sqrt{N}\ket{n, N_b-1,N_f+1},\nn
&&Q^\dagger\ket{n, N_b,N_f}=\sqrt{N}\ket{n, N_b+1,N_f-1}.
\eeq
In fact, since $Q^2=0$, one of these two states is necessarily annihilated by the supercharge, or it{{s}} complex conjugate. Indeed, suppose for instance that $Q^\dagger\ket{n, N_b,N_f}$ exists. By acting on this state with $Q$, one gets $QQ^\dagger \ket{n, N_b,N_f}\propto  \ket{n, N_b,N_f}$. Acting once more with $Q$ yields $Q\ket{n, N_b,N_f}=0$. A similar reasoning can be used to show that if $Q\ket{n, N_b,N_f}$ exists, then it is annihilated by $Q^\dagger$. It follows that any state $\ket{n, N_b,N_f}$ is annihilated either by $Q$ or by $Q^\dagger$. This property has actually been used to determine the coefficient $\sqrt{N}$ in Eq.~(\ref{superNbNf}). Now, 
because $[H,Q]=[H,Q^\dagger]=0$, if $\ket{n, N_b,N_f}$ is an eigenstate of $H$, so are $\ket{n, N_b-1,N_f+1}$ and $\ket{n, N_b+1,N_f-1}$, with the same eigenvalue. However, only two of these three states can simultaneously exist, so that the eigenstates of $H$ are doubly degenerate, except the vacuum state which is invariant under the action of the supercharges $Q$ and $Q^\dagger$. Note that the vacuum state ($N=0$) is the only state that possesses this property, as is clear from Eq.~(\ref{superNbNf}). \\

\begin{figure}[t] 
\begin{center}
\includegraphics[width=0.65\textwidth]{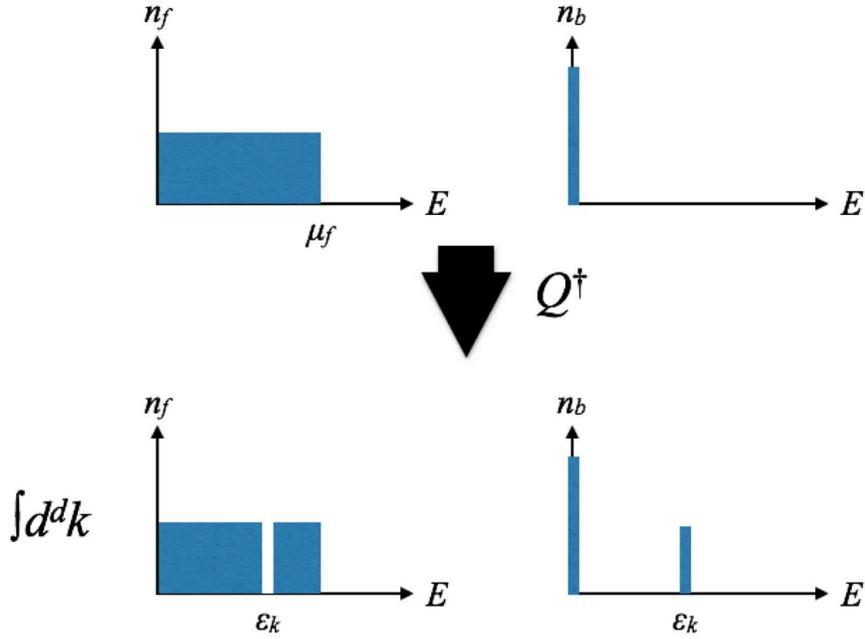}
\caption{{{(Color online)}}
Schematic picture illustrating the particle-hole excitations  (bottom) whose operator is $Q^\dagger$ induced by the action of the operator  $Q^\dagger$ acting on the non interacting ground state (top)
The left (right) panel represents the distribution function of the fermion (boson).}
\label{fig:goldstino-momentum} 
\end{center} 
\end{figure}

In the weak coupling limit,  the ground state will typically be composed of 
 a  Fermi sea of fermions, while the bosons will occupy the zero momentum state, as is schematically shown in the upper panel of Fig.~\ref{fig:goldstino-momentum}. 
 We shall study the spectrum of  long wavelength excitations  induced  in such a system by the action of the supercharge $Q$, or its  complex conjugate $Q^\dagger$, or more generally, by the operators 
 \beq\label{qvp}
q_\vp=\int \rmd^d \vx\, \rme^{-i\vp\cdot\vx} \,q(\vx)=\int \frac{\rmd^d \vk}{(2\pi)^d}\, f^\dagger(\vk) b(\vk+\vp), 
\eeq
and its complex conjugate $q_\vp^\dagger$. The supercharge corresponds to the zero momentum components of $q_\vp$, $Q=q_{\vp=0}$, and reads explicitly
\begin{align} 
Q= \int \frac{\rmd^d \vk}{(2\pi)^d} \,  f^\dagger(\vk)b(\vk),\qquad
Q^\dagger= \int \frac{\rmd^d \vk}{(2\pi)^d}\, b^\dagger(\vk) f(\vk).
\end{align}
At zero temperature, since the bosons are only present in the $\vk=0$ mode,  $Q$ annihilates the ground state as long as $\Nf$ is finite (since then a fermion occupies  the single particle state with  zero momentum). 
By contrast, $Q^\dagger$ generates a superposition of   particle-hole excitations, where the hole is fermionic and the particle is bosonic (see the lower panel of Fig.~\ref{fig:goldstino-momentum}).  These particle-hole excitations are degenerate, a fermion with energy $t_{\rm h}\vk^2$ being simply replaced by a boson with the same energy. This is the situation that we shall mostly consider in this paper.  We shall occasionally comment on the case $N_f=0$, i.e., with no fermion in the ground state. Then,  the action of the supercharge $Q$ will produce an excitation with one fermion in the zero momentum state. Note that in order to avoid having to deal with Bose condensation, we shall restrict ourselves to two-dimensional systems\footnote{It may be needed in some explicit calculations to keep a very small, but non vanishing temperature (and correspondingly a small, negative, chemical potential), in order to stay away from condensation.} ($d=2$). In particular, we assume that the dispersion relation of the bosons remains quadratic at small momenta.
\\

Since we shall be considering excitations where the numbers of bosons and fermions are allowed to change, it is convenient to work in  Fock space, and introduce chemical potentials. We define
\beq
H_G=H_0+V-\mu_b N_b-\mu_f N_f=H-\mu N-\Delta\mu \Delta N, 
\eeq
with $\mu\equiv (\mu_b+\mu_f)/2$, $\Delta N\equiv (N_f-N_b)/2$ and $\Delta\mu\equiv \mu_f-\mu_b$. While the total number of particles $N$ commutes with $H_G$, this is not so of the supercharge $Q$ which does not commute with $\Delta N$. By using the commutation relations, Eqs.~(\ref{algebra}) above,  we get 
 \beq
 [Q, H_G]=\Delta\mu \,Q,\qquad   [Q^\dagger, H_G]=-\Delta\mu \,Q^\dagger.
 \eeq
Thus, the difference of chemical potentials,  $\mu_f-\mu_b=\Delta\mu$,  induces an explicit symmetry breaking that lifts the (twofold) degeneracy between the eigenstates of $H$ (or of $H-\mu N$).  

 In particular, for the state illustrated in Fig.~\ref{fig:goldstino-momentum}, we have
\beq
E_1-E_0=\Delta\mu,
\eeq
where $E_0$ is the energy of the ground state $|\psi_0\rangle$ of $H_G$ (with given number of bosons and fermions) and $E_1$ is the energy of the state $Q^\dagger|\psi_0\rangle$. That is, the particle-hole states have now energies $(t_{\rm h}\vk^2-\mu_b)-(t_{\rm h}\vk^2-\mu_f)=\Delta\mu$ above the ground state (in this case, $\Delta\mu=\mu_f>0$). Note that $\ket{\psi_0}$ and $Q^\dagger|\psi_0\rangle$ are degenerate ground states of $H-\mu N$.  In the case where the ground state contains only bosons, excitations are induced by the supercharge $Q$, and the excitation energy of the state $Q\ket{\psi_0}$ vanishes: $E_1-E_0=-\Delta\mu=\mu_b=0$. In this case, the hamiltonian $H_G$ is supersymmetric (since then $\Delta\mu=0$), and its ground state $\ket{\psi_0}$ is degenerate with the state $\ket{\psi_1}=Q\ket{\psi_0}$. This holds however only for the non interacting system. 
Repulsive interactions yield  positive contributions to 
{{$\mu_f$ and $\mu_b$ (detailed expressions in the mean field approximation are given in Sec.~\ref{sec:spectrum}.), which are in general distinct and result therefore in 
a non vanishing  $\Delta\mu$ and correspondingly an explicit supersymmetry breaking. }}
\\

\subsection{Retarded propagator and excitation spectrum}

Further insight into  the excitation spectrum can be gained by considering the  retarded propagator of the supercharge
\begin{align} 
\label{eq:definition-retarded}
\begin{split}
G^R(t,\vx)&\equiv i\theta(t)\langle \{ q(t,\vx), q^\dagger(0)\}\rangle_{\text{eq}}. 
\end{split}
\end{align} 
At vanishing temperature, the expectation value is a ground state expectation value. We may then write a spectral representation in terms of the excited states $\psi_n$ and $\psi_m$ that can be reached from the ground state by acting respectively with $q^\dagger(\vx)$ and $q(\vx)$. After taking a Fourier transform of the time variable, we obtain
\beq
\label{eq:GR-qq*}
&&G^R(\omega,\vx) =-\left\{ \sum_n \frac{ \bra{\psi_0}q(\vx)\ket{\psi_n}\bra{\psi_n}q^\dagger(\vzero)  \ket{\psi_0}}{\omega-(E_n-E_0)-\Delta\mu} \right.\nn
&&\qquad\qquad \qquad\quad\left. +\sum_m \frac{ \bra{\psi_0}q^\dagger(\vzero)\ket{\psi_m}\bra{\psi_m}q (\vx) \ket{\psi_0} }{\omega+(E_m-E_0)-\Delta\mu }  \right\}.
\eeq
This expression shows that the retarded propagator has poles corresponding to the excitation energies $E_n-E_0$ of the states $\ket{\psi_n}$ that have non vanishing overlap with $q^\dagger(\vx)  \ket{\psi_0}$, and at minus the energies of the states $\ket{\psi_m}$ that overlap with $\sim q(\vx)  \ket{\psi_0}$, the excitation  energies being measured with respect to the chemical potential difference $\Delta\mu$.

By integrating Eq.~(\ref{eq:GR-qq*}) over $\vx$, one obtains the zero momentum component of the retarded propagator, 
\beq
&&G^R(\omega,\vp=\vzero) =-\left\{ \sum_n \frac{ \bra{\psi_0}Q\ket{\psi_n}\bra{\psi_n}q^\dagger(\vzero)  \ket{\psi_0}}{\omega-(E_n-E_0)-\Delta\mu} \right.\nn
&&\quad\qquad\qquad \qquad\quad\left. +\sum_m \frac{ \bra{\psi_0}q^\dagger(\vzero)\ket{\psi_m}\bra{\psi_m}Q \ket{\psi_0} }{\omega+(E_m-E_0)-\Delta\mu }  \right\},
\eeq
Among the intermediate states, there are states that are degenerate with the ground state, and are of the form $Q\ket{\psi_0}$ or $Q^\dagger \ket{\psi_0}$. Note that, for reasons already discussed,  only one of these two states can be realized, and which is realized depends on the sign of $\Delta\mu$. 
If $\Delta\mu>0$, we must have $Q\ket{\psi_0}=0$, otherwise $G^R(\omega,\vp=\vzero)$ has a pole for  $\omega=\Delta\mu$, corresponding to a negative excitation energy $-\Delta\mu$, in contradiction with the assumption that $\ket{\psi_0}$ is the ground state.
Therefore only the first term contributes for $\Delta\mu>0$. By the same reasoning, one shows that  only the second term is finite when $\Delta\mu<0$. Note that this simple property no longer holds at finite momentum $\vp$, as we shall see in Sec.~\ref{sec:spectrum}.

In fact, the states that are degenerate with the ground state entirely dominate the retarded propagator at $\vp=0$.
 To see that, let us return to Eq.~(\ref{eq:definition-retarded}) and write its Fourier transform as follows
\begin{align} 
G^R(\omega,\vp=\vzero)
&= i \int_0^\infty dt \,e^{i\omega t}  
\langle \{Q(t),q^\dagger(0,\vx=\vzero)\}\rangle.
\end{align}
By using $Q(t)=e^{-i\Delta\mu t}Q(0)$, which follows from the equation of motion for the Heisenberg operator $Q(t)$, $i\partial_t Q=[Q,H_G]=\Delta\mu Q$, and the anti-commutation relation 
\begin{align}
\label{eq:canonical-anticommutation}
\{ q(\vx), q^\dagger(\vy)\}&= n(\vx)\delta^d(\vx-\vy),
\end{align}
 which is easily obtained from the (anti-)commutation relation for $f$ and $b$, Eqs.~(\ref{comrel}), we obtain
 \beq
\label{eq:G0-p=0-pole}
 G^R(\omega,\vp=\vzero)= -\frac{\rho}{\omega-\Delta\mu},
\eeq
where $\omega$ is assumed to have a small positive imaginary part to account for the retarded condition. We have set $\rho\equiv \langle n(\vx)\rangle_{\text {eq}}$, the equilibrium density $\langle n(\vx)\rangle_{\text {eq}}$ being assumed uniform, i.e., independent of $\vx$.
 This expression reveals indeed that the retarded propagator has a single pole at $\omega=\Delta\mu$, that can be associated with the state which is reached from the ground state by the action of $Q^\dagger$ or $Q$, depending on whether $\Delta\mu $  is positive or negative, as we have just discussed.  
 
 This result was obtained using only symmetry considerations and commutation relations. It is independent of the details of the model Hamiltonian, such as the strength of the interaction, and it holds at finite temperature as well~\cite{Nicolis:2012vf,Watanabe:2013uya,Hayata:2014yga}. 
 The pattern we observe here is  reminiscent  of the standard pattern of spontaneous symmetry breaking \cite{Nambu:1961tp}, with an underlying degeneracy that yields a mode with vanishing excitation energy in the absence of explicit symmetry breaking:  the location of the pole in Eq.~(\ref{eq:G0-p=0-pole})  goes to zero as the explicit symmetry breaking term $\Delta\mu$ vanishes. Because of this analogy with the Nambu-Goldstone excitations, this mode is generally referred to as  the ``Goldstino''.  It is a collective excitation, as reflected for instance in the fact that  the spectral  strength is proportional to the total density $\rho$. But in contrast to most collective excitations of many-body systems, which are bosonic in nature, the Goldstino carries fermionic quantum number. 
The total density $\rho=\rho_f+\rho_b$ plays the role of an order parameter for SUSY breaking (recall that the action of the supercharge on a given state is proportional to the total number of particle in that state, see Eq.~(\ref{superNbNf})). Note that this order parameter is independent from the explicit breaking induced by the difference of chemical potentials, and hence the difference in the densities $\rho_f-\rho_b$.
 
\subsection{Spectral sum rules}
\label{sec:sum-rule}
The spectral function  $\sigma(\omega,\vp)=2{\text{Im}}G^R(\omega,\vp)$: 
\beq
\sigma(\omega,\vp)=\int_{-\infty}^{\infty} \rmd t \,\rme^{i\omega t}\int d^2\vx \,e^{-i\vp\cdot \vx}
\langle \{ q(\vx,t),q^\dagger(0)\}\rangle
\eeq
obeys sum rules which can be used to obtain model independent information on the properties of the Goldstino at finite momentum. We shall consider here more specifically two sum rules. 
The first one reads
 \begin{align}
 \label{eq:sum-rule}
\begin{split}
\int^\infty_{-\infty} \frac{d\omega}{2\pi}\sigma(\omega,\vp)
&= \int d^d\vx\, e^{-i\vp\cdot\vx} 
\langle\{q(0,\vx), q^\dagger(0,\vzero)\} \rangle=\rho,
\end{split}
\end{align}
where in the last step, we have used the (anti)-commutation relations {{Eq.~}}(\ref{comrel}). 
The value of this sum rule coincides with the residue of the pole at $\omega=\Delta\mu$ in Eq.~(\ref{eq:G0-p=0-pole}). 
In other words,  the Goldstino exhausts the sum rule at $|\vp|=0$, as we have already observed. 
{{We also emphasize that this sum rule is valid regardless of detail of the Hamiltonian or temperature, since we have used only the canonical (anti)-commutation relations in its derivation.
}}

The second sum rule follows by  noticing that the supersymmetry entails a local conservation law. To see that, consider the equation of motion $i\del_t q(t,\vx)=[q(t,\vx),H_G]$. A simple calculation yields
\beq
\del_t q(t,\vx)=i t_{\rm h}  \left[ f^\dagger(\vx) \nabla^2 b(\vx)-\left(  \nabla^2 f^\dagger(\vx) \right) b(\vx)   \right]
-i\Delta\mu q(\vx), 
\eeq
which we can put in the form
\beq\label{contQ}
\del_t q(t,\vx)+\nabla\cdot \vj_Q=-i\Delta\mu \,q(t,\vx),
\eeq
with the current
\beq
\vj_Q(\vx)=\frac{t_{\rm h}}{i}\left[ f^\dagger(\vx) \nabla b(\vx)-  \left(  \nabla f^\dagger(\vx) \right) b(\vx) \right].
\eeq
This current is the analog, for the supercharge, of the particle currents for bosons and fermions, e.g., 
\beq
\vj_{b}=\frac{t_{\rm h}}{i}\left[ b^\dagger(\vx) \nabla b(\vx)-  \left(  \nabla b^\dagger(\vx) \right) b(\vx) \right],
\eeq
and similarly for $\vj_f$. These currents obey continuity equations expressing the local conservation of $N_b$ and $N_f$, e.g., 
\beq
\del_t n_b(t,\vx)+\nabla\cdot \vj_b=0, \qquad \del_t n_f(t,\vx)+\nabla\cdot \vj_f=0.
\eeq
Equation (\ref{contQ}) has the form a continuity equation for the supercharge, with a source term proportional to the explicit symmetry breaking term. 
Taking the Fourier transform of this equation, we obtain
\beq
\label{eq:EOM-supercharge-p}
i\del_t q_\vp(t)=t_{\rm h}\int_{\vk}\vk^2 \left(f^\dagger_{\vk-\vp} b_{\vk} - f^\dagger_{\vk} b_{\vk+\vp} \right)+\Delta\mu \,q_\vp ,
\eeq
with $q_\vp(t)$ given by Eq.~(\ref{qvp}). 
We note that the first term in the right-hand side of Eq.~(\ref{eq:EOM-supercharge-p}) vanishes at $|\vp|=0$, leaving the equation of motion $i\del_t Q(t)=\Delta\mu Q$ which we have used earlier, with the identification $Q=q_{\vp=0}$. 

We can use Eq.~(\ref{eq:EOM-supercharge-p}) in order to evaluate a second sum rule
\beq
\int_{-\infty}^{\infty}\frac{\rmd\omega}{2\pi} \,\omega\sigma(\omega,\vp)
&=&\int_\vq \left.\langle \{ i\del_t q(\vp,t),q^\dagger(\vq)\}\rangle\right|_{t=0}.
\eeq
We have
\beq
\int_{-\infty}^{\infty}\frac{\rmd\omega}{2\pi} \,\omega\sigma(\omega,\vp)&=&\rho\Delta\mu
+t_{\rm h}\int_\vq \int_{\vk}\vk^2
\langle \{ \left(f^\dagger_{\vk-\vp} b_{\vk} - f^\dagger_{\vk} b_{\vk+\vp} \right),q^\dagger(\vq)\}\rangle\nn
&=&\rho\Delta\mu
+t_{\rm h}\vp^2 \int_{\vk}
\langle f^\dagger_{\vk} f_{\vk} - b^\dagger_{\vk} b_{\vk} \rangle \nn
\label{eq:sumrule-2nd}
&=&\rho\Delta\mu
+t_{\rm h}\vp^2 (\rhof-\rhob),
\eeq
where we have used the (anti)commutation relations.
{{
This sum-rule, analogous to the f-sum rule, was derived without any approximation, by using the form of the supercurrent and the canonical (anti)commutation relations. It  involves only the kinetic part of the hamiltonian that we have considered, and is therefore independent of the interaction strength $U$. It could be modified in case non local (momentum dependent) interactions are present. 
}}

For $\vp$ not too big, we may expect the Goldstino to continue to exhaust the sum rule, with a single peak in the spectral function located at $\omega=\omega_{\rm s}(p)$, with residue $\rho$.  When this is the case, the left hand side equals $\omega_{\rm s} \rho$ and the sum rule (\ref{eq:sumrule-2nd}) yields
\beq
\label{eq:result-dispersion-assumption}
\omega_{\rm s}(p)
&=&\Delta\mu
-\alpha_{\rm s} \vp^2,
\eeq
where 
\begin{align}
\label{eq:alpha-expression-largeU}
\alpha_{\rm s}&\equiv \frac{t_{\rm h}}{\rho}(\rhob-\rhof). 
\end{align}
We recover the energy of the Goldstino for $p=0$, $\omega_{\rm s}(p)=\Delta\mu$. In addition, the sum rule provides an explicit expression for  the  dispersion relation. This quadratic dispersion relation coincides with that obtained from a construction of a simple effective action based on supersymmetry~\cite{Bradlyn:2015kca}.
However, the  assumption that the Goldstino exhausts the sum rule at finite momentum is not realized in generic situations with a finite fermion density,  as the detailed analysis of the next section will show.
Nevertheless, the sum rules derived in this section remain useful even in such cases, as checks of our numerical calculations.

\section{Explicit calculation within the Random Phase Approximation}
\label{sec:spectrum}

We turn now to an explicit calculation of the Goldstino properties using the RPA. 
As mentioned earlier, we specialize to two dimensions.
This is  in line with the possible realization of the phenomenon suggested in Ref.~\cite{Shi:2009ak}. 
Also, working in two dimensions allows us to avoid dealing with  Bose-Einstein condensation, which would introduce new features beyond the scope of the present paper.  

\begin{figure}[t] 
\begin{center} 
\includegraphics[width=0.7\textwidth]{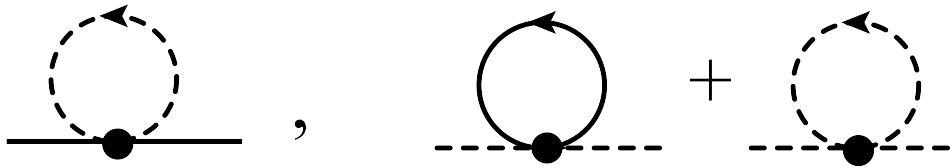}
\caption{The self-energies for the fermion (left panel) and the boson (right panel) in the mean field approximation. The full (dashed) line represents a fermion (boson) propagator.
}
\label{fig:one-loop}
\end{center}
\end{figure}


The generic situation that we consider is that in which the fermion density is finite, given  by  $\rho_f=k_F^2/(4\pi)$, with $k_F$ the Fermi momentum. 
In the absence of interactions the fermion chemical potential is $\mu_f=\epsilon_F$, with $\epsilon_F=t_{\rm h}k_F^2$ the Fermi energy. The boson chemical potential is 
$\mu_b=0$  at $T=0$, and thus $\Delta\mu^0=\epsilon_F$ is positive, where $\Delta\mu^0$ is the value of $\Delta\mu$ for $U=0$.

In the presence of interactions, the  chemical potentials $\muf$ and $\mub$ are determined by the constraints $\rhof= \int_\vk n^f_\vk$, $\rhob= \int_\vk n^b_\vk$,  
where $n^f_\vk\equiv1/(\exp({ (\epsilon^f_\vk-\mu_f)/T})+1)$, $n^b_\vk\equiv 1/(\exp({ (\epsilon^b_\vk-\mu_b)/T})-1)$ are the Fermi and Bose statistical factors,  and $\int_\vk$ is a shorthand for $\int {\rmd^2 \vk}/{(2\pi)^2}$. In the mean field approximation, the single particle energies are given by (see Fig.~\ref{fig:one-loop} for the relevant diagrams)
\beq\label{spenergies}
\epsilon^f_\vk=t_{\rm h} \vk^2+U\rho_b,
\qquad \epsilon^b_\vk=t_{\rm h} \vk^2+U(2\rho_b+\rho_f).
\eeq
At zero temperature the chemical potential are then given in terms of the densities by
$
\mu_f=\epsilon_F+U\rho_b$,  $\mu_b=U(2\rho_b+\rho_f),
$
so that $\Delta\mu=\mu_f-\mu_b=\epsilon_F-U\rho$, with $\rho=\rho_f+\rho_b$. 
These values of the chemical potentials are those which maintain the densities equal to their values in the absence of interaction. 
Note that, as we increase the value of $U$ from zero, keeping the densities fixed,  $\Delta\mu$ changes its sign from positive to negative at $U=4\pi t_{\rm h} \rho_f/\rho$, at which point the nature of the Goldstino excitation changes (see the discussion in Sec.~\ref{sec:Intro-2}).

At finite temperature, the chemical potentials are determined as functions of the fermion and boson densities from the following equations
\beq
&&\muf=U\rho_b+T\ln(e^{4\pi t_{\rm h}\rho_f/T}-1),\nn
&&\mub=U(2\rhob+\rhof)+ T\ln(1-e^{-4\pi t_{\rm h}\rho_b/T}).
\eeq 
In fact, within the range of parameters that we shall be using, the chemical potentials will never differ much from their zero temperature values. 

We are using units with $\hbar=1$. Then, if $a$ denotes the unit length, a momentum has dimension  $[p]=a^{-1}$. The quantity $t_{\rm h}p^2$ has dimension of an energy, i.e.,  $[t_{\rm h}]=[E] a^2$. Energies will then be measured as ratios $t_{\rm h}a^{-2}$. The ratio $U/t_{\rm h}$ is dimensionless. By choosing the unit length equal to unity, we make the dimension of $t_{\rm h}$ that of an energy. We shall use this quantity as energy unit. 
We are also using $k_B=1$, so that the temperature has the same unit as the energy. We shall consider temperatures in the range $0\lesssim T\lesssim t_{\rm h}$.

In the numerical calculations presented in this section, we have made a specific choice for the densities, $\rhof=0.5$ and $\rhob=1$. This is in order to illustrate the generic situations that we wish to discuss, avoiding in particular special values (e.g if $\rhof=\rhob$, the right hand side of the sum rule Eq.~(\ref{eq:sumrule-2nd}) vanishes).


\begin{figure}[t] 
\begin{center}
\includegraphics[width=0.5\textwidth]{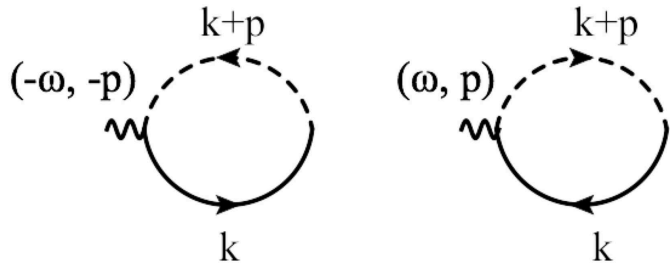}
\caption{The retarded Green function in free limit, $G^0(p)$.
These diagrams represent particle hole excitations:
The left panel is for fermion particle-boson hole, while the right one is for fermion hole-boson particle.}
\label{fig:G0} 
\end{center} 
\end{figure} 

\subsection{Free case ($U=0$)}
\label{ssc:U=0}

At $U=0$, the retarded Green function in the supercharge channel is given by the one-loop diagrams displayed in Fig.~\ref{fig:G0}. It reads (recall that $\omega$ has a small (positive) imaginary part)
\beq
\label{eq:goldstino-one-loop_a}
G^0(\omega, \vp)
= -\int \frac{d^2\vk}{(2\pi)^2}\left\{ \frac{n^f_\vk(1+n^b_{\vk+\vp})}{\omega -(\epsilon^b_{\vk+\vp}-\ef+\Delta\mu^0)}+\frac{n^b_{\vk+\vp}(1-n^f_\vk)}{\omega +(\ef-\epsilon^b_{\vk+\vp}-\Delta\mu^0)}\right\}.\nn
\eeq
This expression makes explicit the origin of the different contributions, as well as their physical interpretation. The two terms are in one-to-one correspondence with the corresponding terms of Eq.~(\ref{eq:GR-qq*}): the first term has poles at positive $\omega$ associated with fermion hole excitations, while the poles of the second term represent (minus) fermion particle excitation energies at negative $\omega$.  
At small $T$, in each term the Bose occupation factor operates a severe selection of available momenta, either by amplifying the scattering at low momentum (in the first term), or by restricting the momenta of possible boson hole states (second term). This restriction of the phase space leads to sharp peaks in the spectral function, as we shall see soon. Note that, in agreement with the arguments developed in the previous section, only one term of Eq.~(\ref{eq:goldstino-one-loop_a}) contributes at $\vp=0$. If $\rho_f\ne 0$, the last term vanishes: $|\vk|$ cannot be both greater than $k_F$ (because of the factor $(1-n^f_\vk)$) and vanish (because of the factor $n^b_{\vk+\vp}=n^b_{\vk}$). If $\rho_f=0$, it is the first term that vanishes. 

\begin{figure}[!t] 
\begin{center} 
\includegraphics[width=0.7\textwidth]{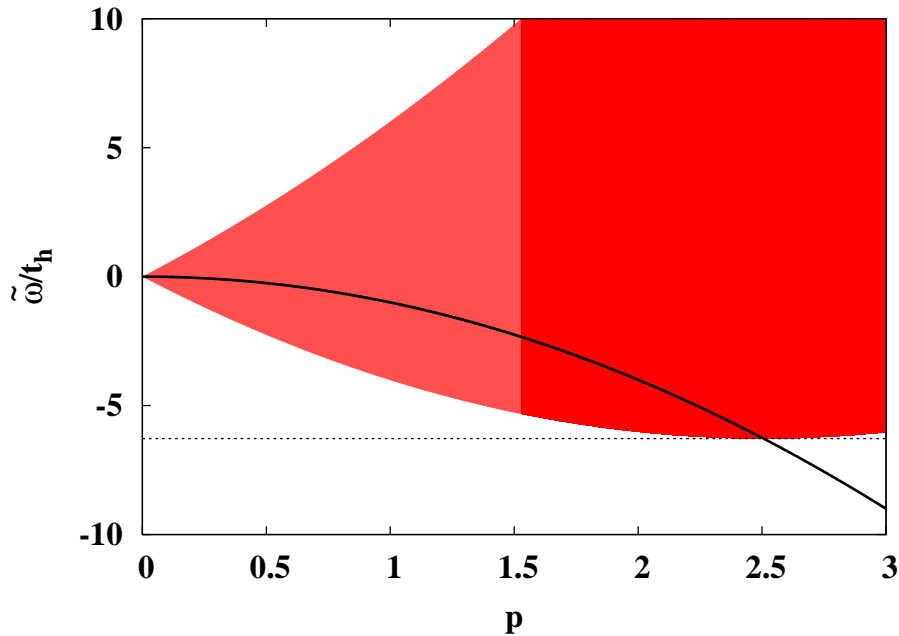}
\caption{{{(Color online)}}
The range of the support for the fermion hole continuum is indicated by the red shaded area. 
The black curve indicates the location of the boson particle pole position (for $p<k_F$) and the boson hole pole for $p>k_F$. 
The parameters are,  $T=U=0$, $\rho_b=1$, $\rho_f=0.5$, corresponding to $k_F=\sqrt{2\pi}\simeq 2.5$.  The dashed horizontal line indicates the value $\tilde\omega=-\Delta\mu^0$, corresponding to $\omega=0$. 
It touches the minimum of the continuum at $p=k_F$. When $p>k_F$, the boson hole pole lies outside the fermion continuum, and its energy $\omega<0$ corresponds to minus the excitation energy of the corresponding Goldstino.
{{We have chosen $a=1$, so that both $\tilde{\omega}/t_h$ and $p$ are dimensionless (see the discussion at the beginning of Sect.~\ref{sec:spectrum}).
}}
} 
\label{fig:support-pole-free}
\end{center}
\end{figure}

At this point, we notice that the denominators in the two terms are in fact identical, so that simplifications occur, leading after some calculation to the more compact expression
\begin{align}
\label{eq:goldstino-one-loop}
\begin{split}
G^0(\omega, \vp)
&= -\int \frac{d^2\vk}{(2\pi)^2} \frac{\nf(\ef)+\nb(\epsilon^b_{\vk+\vp})}{\omega -\epsilon^b_{\vk+\vp}+\ef-\Delta\mu^0}\\
&=  -\frac{1}{4\pi t_{\rm h}|\vp|}  \sum_{i=f,b}
\Biggl(
\int^{k_{ci}}_{0} d|\vk| |\vk| \frac{n_i(\epsilon^i_\vk)\sgn(\tilde{\omega}_i)}{\sqrt{k^2_{ci}-|\vk|^2}} \\
&~~~ -i \int^{\infty}_{k_{ci}} d|\vk| |\vk| 
\frac{n_i(\epsilon^i_\vk)}{\sqrt{|\vk|^2-k^2_{ci}}} \Biggr),
\end{split}
\end{align}
where $\tilde\omega\equiv\omega-\Delta\mu^0$, $\tilde{\omega}_f\equiv \tilde{\omega}-t_{\rm h}\vp^2$, $\tilde{\omega}_b\equiv \tilde{\omega}+t_{\rm h}\vp^2$, $k_{cf}\equiv |\tilde{\omega}-t_{\rm h}\vp^2|/(2t_{\rm h}|\vp|)$ and $k_{cb}\equiv  |\tilde{\omega}+t_{\rm h}\vp^2|/(2t_{\rm h}|\vp|)$.    
At low temperature,  $\mu_f\simeq \epsilon_F$ and $\mu_b\simeq 0$, so that $\Delta\mu^0\simeq \epsilon_F$.
We see from Eq.~(\ref{eq:goldstino-one-loop}) that the imaginary part of $G^0$ comes from the region of $\omega$ values where the equation 
\beq\label{phenergies}
\omega=\epsilon^b_{\vk+\vp}-\epsilon^f_\vk+\Delta\mu^0=\Delta\mu^0+t_{\rm h}(\vp^2+2\vp\cdot\vk),
\eeq
is satisfied. Note that the minimum of the continuum occurs for $p=k_F$: for this value of the momentum, there exists an excitation where a fermion on the top of the Fermi sea ($k=k_F$) turns into a boson with vanishing momentum, releasing an energy $\epsilon_F$  that just compensates $\Delta\mu^0$. This particular excitation occurs therefore at $\omega=0$. 
It is degenerate with the reverse process where a boson receives a kick of momentum $k_F$ and turns into a fermion at the top of the Fermi sea. This degeneracy is visible in Fig.~\ref{fig:support-pole-free}: it is the point where the branch corresponding to the boson hole excitation leaves the fermion continuum, at $p=k_F$. 

At $|\vp|=0$, $G^0(\omega, \vp=0)$ has a pole at $\omega=\Delta\mu^0\simeq\epsilon_F$, corresponding to the free Goldstino excitation. Since $\Delta\mu^0>0$, this excitation is realized by the action of the operator $Q^\dagger$ on the ground state. 

As $\vp$ increases, the pole turns into a branch cut singularity corresponding to a continuum of boson particle -fermion hole  excitations with energies given by Eq.~(\ref{phenergies}).
The corresponding branch cut has support in the region 
\beq
t_{\rm h}p^2-2t_{\rm h}k_Fp\le \tilde{\omega} \le t_{\rm h}p^2+2t_{\rm h}k_Fp.
\eeq 
In addition,  there is another contribution whose phase space is controlled by the Bose statistical factor, which imposes $\vk+\vp\simeq 0$. This leads to a delta function contribution to the spectral function located at 
 \begin{align}
\label{eq:dispersion-largep}
\omega=\Delta\mu^0 -t_{\rm h}|\vp|^2.
\end{align}
 This delta peak is located within the fermion continuum as long as $p<k_F$, and moves out of it when $p>k_F$. 
This behavior is illustrated in Fig.~\ref{fig:support-pole-free}.
The physical interpretation of the mode changes depending on whether $p<k_F$ or $p>k_F$. In the former case, we are dealing with  a boson particle (fermion hole) state, while in the latter we are dealing with a boson hole (fermion particle) state (see the discussion after Eq.~(\ref{eq:goldstino-one-loop_a}) above).  

\begin{figure}[t] 
\begin{center} 
\includegraphics[width=0.5\textwidth]{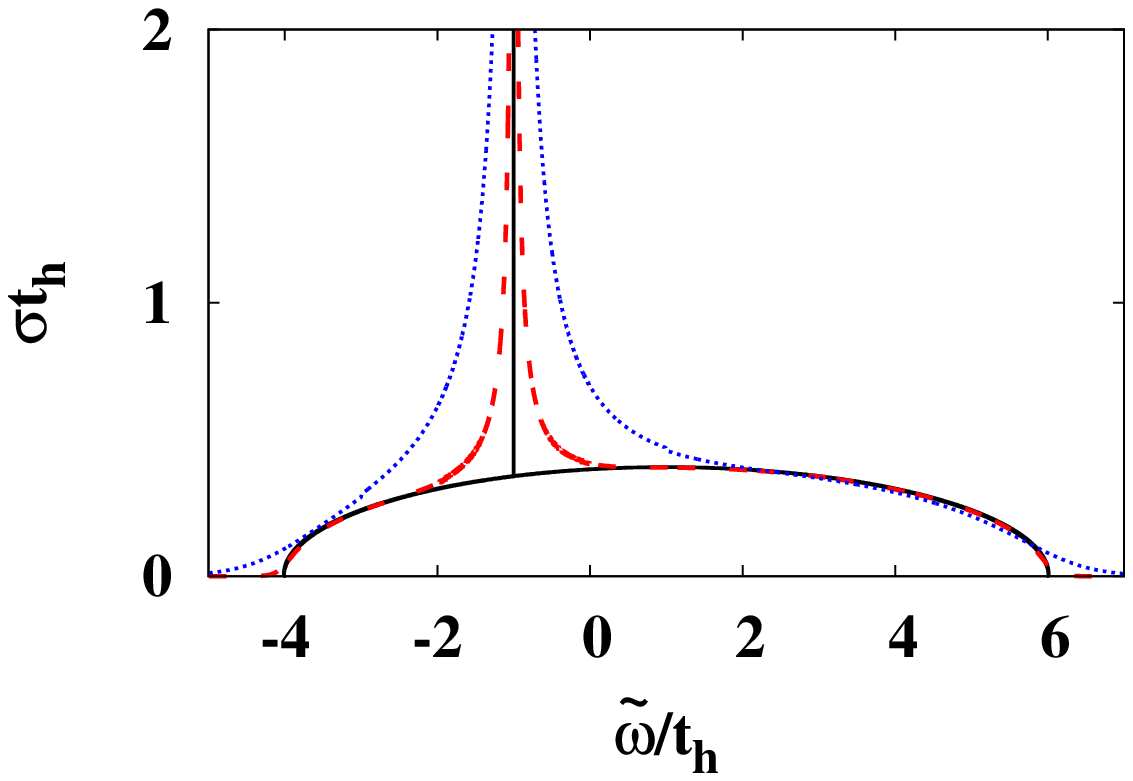}\includegraphics[width=0.5\textwidth]{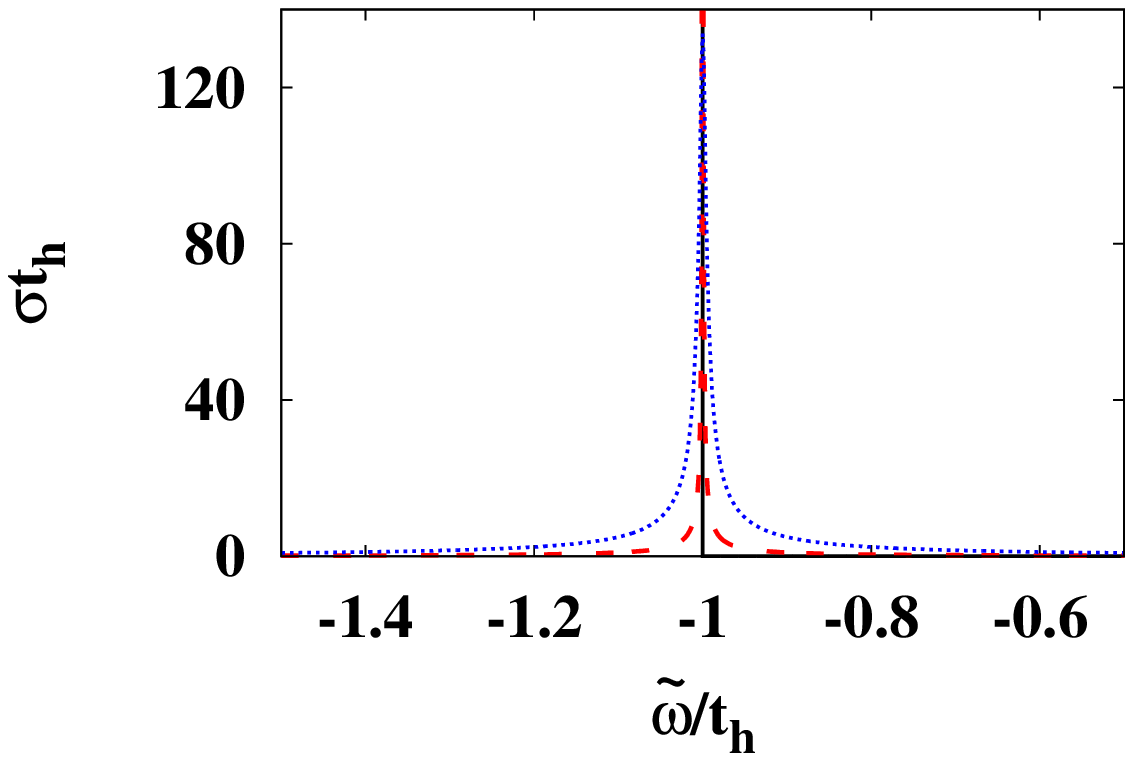}
\caption{{{(Color online)}}
 The free ($U/t_{\rm h}=0$) spectral function at $|\vp|=1.0$.
The solid (black), dashed (red), and dotted (blue) lines correspond to temperatures $T/t_{\rm h}=0$, $0.2$, and $1.0$, respectively. 
{{The unit of $\sigma$ is $1/t_h$.
}}
Left panel: excitations induced by $q^\dagger_\vp$. The large width  is due to the fermion continuum ($\rho_f=0.5, \rho_b=1$). 
Right panel: 
excitations induced by $q_\vp$, with vanishing fermion density ($\rhof=0, \rhob=1$).
} 
\label{fig:spectrum-free}
\end{center}
\end{figure}

The spectral function at $T=0$ is given by the following explicit expression 
 \begin{align}
\label{eq:spectrum-T=0-largep}
\sigma(\omega,\vp)&= \frac{\theta(k_F-k_{cf})}{2\pi t_{\rm h}|\vp|}\sqrt{k^2_F-k^2_{cf}}
+2\pi \rhob\delta(\tilde{\omega}+t_{\rm h}|\vp|^2).
\end{align}
It is plotted as a function of $\tilde{\omega}$ in Fig.~\ref{fig:spectrum-free}, left panel, for the value of the momentum $|\vp|=1.0$. The integrals of the spectral weights of the continuum and the peak are respectively $\rho_f $ and $\rho_b$. They add up to $\rho=\rho_f+\rho_b$, in accordance with the sum rule, Eq.~(\ref{eq:sum-rule}). It is also easily checked  that Eq.~(\ref{eq:spectrum-T=0-largep}) satisfies the second sum rule, Eq.~(\ref{eq:sumrule-2nd}). 

  One can also verify that when $p\ll  k_F$, the fermionic continuum goes over to a delta function $2\pi \rho_f \delta(\tilde{\omega}-t_{\rm h}p^2)$. 
  Nevertheless, there is no momentum region where the spectral function exhibits two separate peaks: 
  The width of the continuum, $4t_{\rm h}|\vp|k_F$,  always exceeds the separation $2t_{\rm h}\vp^2$ between such possible peaks.

  It is worth emphasizing in this regard the large effect of the Fermi sea. In ordinary fermion systems, long wavelength excitations involve particle-hole excitations forced by the Pauli principle to be located in the immediate vicinity of the Fermi surface. 
  As a result the widths of the collective excitations vanish at low momentum. In the present case, particle-hole states involving a boson as particle are not affected by the Pauli principle. As a result, the entire Fermi sea, and not only the vicinity of the Fermi surface contributes to the excitations, yielding a huge phase space, which is responsible for the large broadening of the excitation as soon as $p$ increases. This can be contrasted with the situation where the Fermi sea is absent. Then a single peak remains, that associated with  the bosonic (hole) excitation (see Fig.~\ref{fig:spectrum-free}, right panel, where $\rhof$ is set to be zero). 
  
Finite temperature contributes  a small width to the latter excitation, while its effect  is larger on the fermion hole excitation, as can be seen in the plots for $T/t_{\rm h}=0.2$ and $T/t_{\rm h}=1.0$ in Fig.~\ref{fig:spectrum-free}. However, in either case, the qualitative picture remains unaltered as the temperature increases, at least as long as it remains moderate, i.e. as long as $T\leq t_{\rm h}$. 


\subsection{The Goldstino at $\vp=0$}

We now consider the effect of the interactions. In the RPA, the single particle energies are evaluated in the mean field approximation, where the effect of the boson-fermion or boson-boson interactions is to shift the single particle energies by a constant amount, $U\rho_b$ for the fermions, and $U(\rho_f+2\rho_b)$ for the bosons (see Eq.~(\ref{spenergies})).  The free response at zero momentum becomes then
\begin{align}
\label{eq:p=0-pole}
G^0(\omega,\vp=\vzero)
&= -\frac{\rho}{\omega-\Delta\mu-U\rho}.
\end{align}

\begin{figure}[t] 
\begin{center} 
\includegraphics[width=0.6\textwidth]{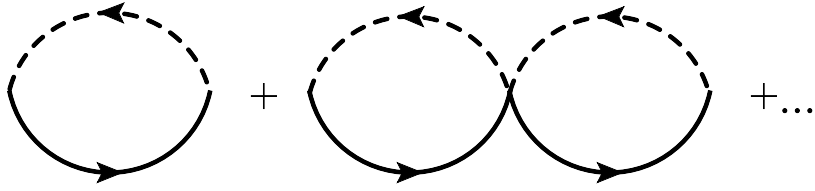}
\caption{`Bubble' diagrams contributing to the Goldstino propagator in the RPA. The diagrams drawn correspond to fermion-particle (full line) boson-hole (dashed line) excitations. There exist 
similar diagrams, not drawn,  with fermion-hole and boson-particle, or mixtures of the various processes. 
} 
\label{fig:RPA}
\end{center}
\end{figure}

After summing the bubble diagrams drawn in Fig.~\ref{fig:RPA}, the full retarded propagator is obtained as 
\beq
\label{eq:RPA-p=0}
G^R(\omega)=\frac{G_0(\omega)}{1-UG_0(\omega)}.
\eeq
Note that for $\omega\simeq \Delta\mu$, $G^0$ is of order $U^{-1}$ from Eq.~(\ref{eq:p=0-pole}), and the summation of  all the bubble diagrams is needed for consistency: Each new bubble brings a factor  $G^0 U\sim U^{0}$, where $U$ comes from the vertex.
 It is easy to verify that, Eq.~(\ref{eq:RPA-p=0}) is identical to Eq.~(\ref{eq:G0-p=0-pole}): there is a complete cancellation of the interaction effects between the corrections to the self energies, and the direct particle-hole interaction. This is a pattern familiar in the study of collective modes associated with spontaneously broken symmetries~\cite{Nambu:1961tp}, and indeed the existence  of a pole at $\omega=\Delta\mu$ is guaranteed by symmetry, as we discussed in the previous section. 

\subsection{The Goldstino at finite momentum}

\begin{figure}[t] 
\begin{center} 
\includegraphics[width=1.03\textwidth]{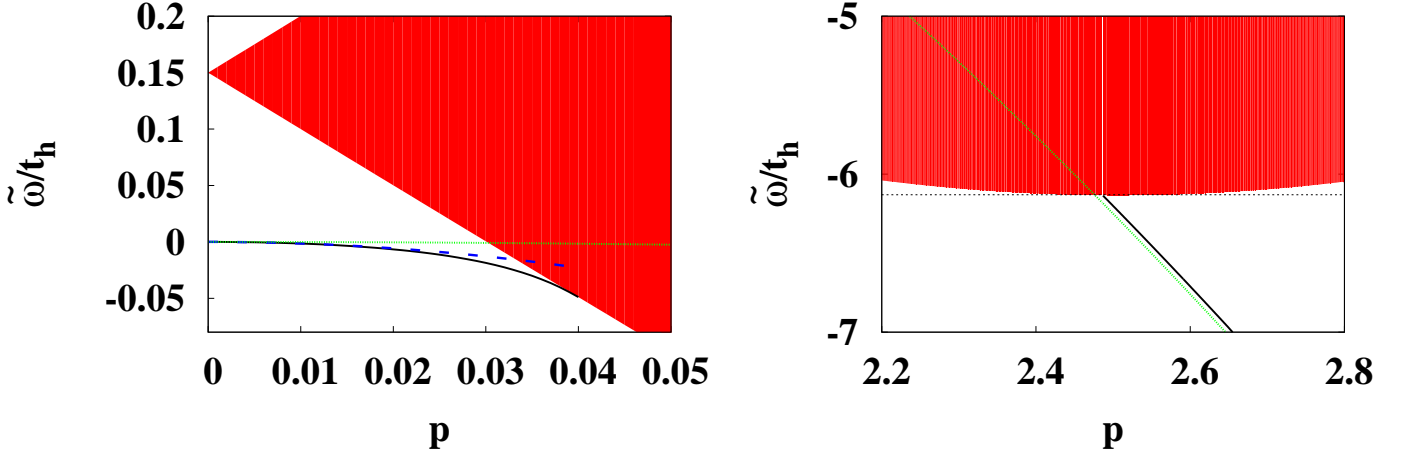}
\caption{{{(Color online)}}
 The fermion continuum (red shaded area) and the position of the Goldstino pole (solid black line) at small $|\vp|$ (left panel) and large $|\vp|$ (right panel). At small $|\vp|$ the quadratic dispersion relation of the Goldstino is given by  $\tilde{\omega}=-\alpha|\vp|^2$, with $\alpha$  well approximated by the expansion (\ref{eq:alpha-expression}). This is indicated by the dashed blue line, while the quadratic dispersion at large $|\vp|$, given by $\tilde{\omega}=-t_{\rm h}|\vp|^2$, is plotted as the dotted green line (the nearly horizontal line in the left panel).
The dashed horizontal line in the right panel indicates the value $\tilde\omega=-\Delta\mu$, corresponding to $\omega=0$. 
 The interaction strength is  $U/t_{\rm h}=0.1$.
} 
\label{fig:support-pole}
\end{center}
\end{figure}

At finite $\vp$, the retarded propagator is given by an equation similar to Eq.~(\ref{eq:RPA-p=0}) and reads
 \begin{align}
\label{eq:result-RPA}
G^R(\omega,\vp)&= \frac{G_0(\omega,\vp)}{1-UG_0(\omega,\vp)},
\end{align}
where $G_0(\omega,\vp)$ is given by Eq.~(\ref{eq:goldstino-one-loop}) with mean field corrections to single particle energies included.

As  illustrated in the left panel of Fig.~\ref{fig:support-pole}, the continuum is shifted up by the constant amount $U\rho$, without any other alteration. In particular, its minimum remains located at momentum $p=k_F$. Because of the shift,  the pole is now located outside the continuum. It continues to exist over a small range of momenta, until the corresponding dispersion relation hits the continuum (at $|\vp|\approx 0.04$ for the present values of the parameters). 
In this range of momenta, we determine numerically the dispersion relation by solving the equation $UG_0(\omega_p,\vp)=1.$  In the vicinity of  $\omega_p$, we expand the propagator, and get
\beq
G(\omega,\vp)\approx -\frac{Z}{\omega-\omega_p},\qquad 
Z=\frac{1}{U^2} \left(\left.\frac{\del G_0}{\del\omega} \right|_{\omega_p} \right)^{-1}.
\eeq
We used this expression to evaluate the residue numerically.

At small momentum, we may also obtain a simple analytical estimate. To that aim, we first expand $G^0$ in terms of $|\vp|$: 
\begin{align}
\label{eq:smallp-expansion}
G^0= -\frac{\rho}{\tilde{\omega}-U\rho}+A\frac{\vp^2}{(\tilde{\omega}-U\rho)^2}
+B\frac{\vp^2}{(\tilde{\omega}-U\rho)^3}
+{\cal O}(\vp^4),
\end{align}
 where the terms that are proportional to $|\vp|$ or $|\vp|^3$ vanish due to the rotational symmetry.
The coefficients $A$ and $B$ are given by
\begin{align} 
A&= t_{\rm h}(\rhob-\rhof),\\
B&= -\frac{t_{\rm h}^2}{\pi}\int^\infty_0 d |\vk| |\vk|^3 (\nf(\ef)+\nb(\eb)). 
\end{align}
Next, we expand Eq.~(\ref{eq:smallp-expansion}) for small $\tilde{\omega}$, which yields
\begin{align}\label{eq:smallp-expansion2}
G^0&=  \frac{1}{U}\left[1+\frac{\tilde{\omega}}{U\rho}+\left(\frac{\tilde{\omega}}{U\rho}\right)^2+\frac{\vp^2}{U\rho^2}\left(A-\frac{B}{U\rho}\right)
+\frac{\tilde{\omega}\vp^2}{U^2\rho^3}\left(2A-\frac{3B}{U\rho}\right)\right]\notag\\
&\quad+{\cal O}(\tilde{\omega}^3,\tilde{\omega}^2\vp^2,\vp^4).
\end{align}
At small temperature, and in the vicinity of $\tilde{\omega}=0$, the $A$ term in Eq.~(\ref{eq:smallp-expansion})    is of order $t_{\rm h}\vp^2/(U^2\rho)$ and the $B$ term is of order $t_{\rm h}^2\vp^2 k_f^4/(U^3\rho^3)\sim t_{\rm h}^2\vp^2/(U^3\rho)$, while the  first term is of order $1/U$. 
At weak coupling $U\ll t_{\rm h}$, the $A$ term is therefore much smaller than the $B$ term.
\begin{figure}[t] 
\begin{center} 
\includegraphics[width=1.0\textwidth]{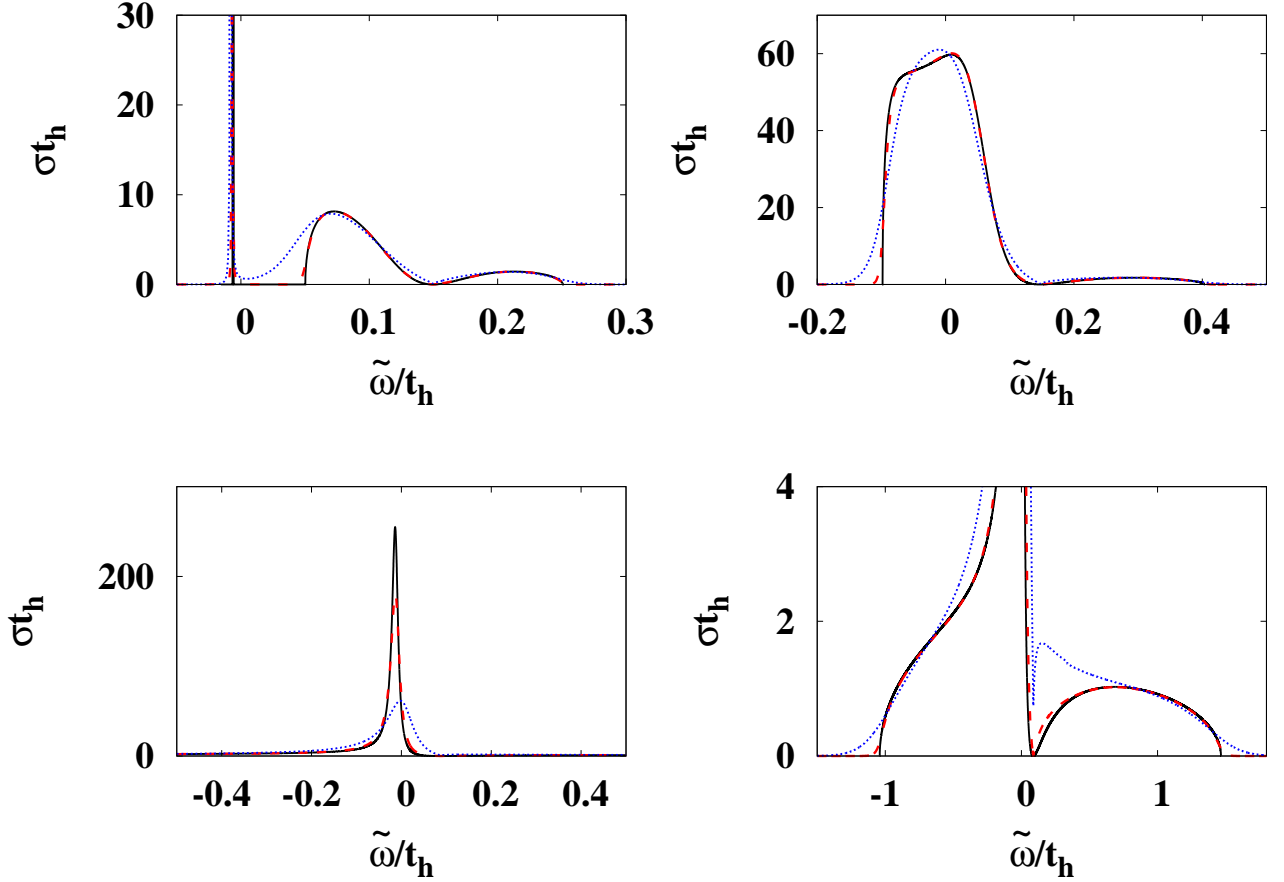}
\caption{{{(Color online)}}
 The spectral function at various temperature.
The results at $|\vp|=0.02$, 0,05, and 0.25 are plotted in the upper-left (0.002), upper-right (0.05), and lower panels (0.25).
The large (small) $\sigma$ part is plotted in the lower-left (lower-right) panel so that the shape of the peak (continuum) becomes clearer.
The solid (black) line corresponds to $T/t_{\rm h}=0$, the dashed (red) line to $T/t_{\rm h}=0.2 $, and the dotted (blue) line to $T/t_{\rm h}=1$, respectively. The plot on the lower-right panel is a blow-up of the lower left panel, to show the structure of the peak near $\tilde\omega=0$. 
} 
\label{fig:spectrum-T=0}
\end{center}
\end{figure}

By using the expansion (\ref{eq:smallp-expansion2}) in  Eq.~(\ref{eq:result-RPA}), we obtain the retarded propagator in the form
\beq 
\label{eq:GR-result-linear-response}
G^R(\omega,\vp)\simeq -\frac{Z}{\tilde{\omega} +\alpha \vp^2 }.
\eeq
Here $\alpha$ and $Z$ are given respectively by
\beq
\label{eq:alpha-expression}
\alpha&\equiv& \frac{1}{\rho}\left(A-\frac{B}{U\rho}\right) \nn
\nonumber
&=& \frac{t_{\rm h}(\rhob-\rhof)}{\rho} 
+\frac{t_{\rm h}^2}{\pi U\rho^2}\int^\infty_0 d |\vk| |\vk|^3 (\nf(\ef)+\nb(\eb)),\nn
 &\approx& \frac{t_{\rm h}}{\rho}
\left\{(\rhob-\rhof)+ \frac{4\pi t_{\rm h}\rhof^2}{U\rho}\right\}. 
\eeq
and
\beq\label{eq:residue-expression}
Z&\equiv& \rho\left(1+\frac{B}{U^2\rho^3}\vp^2\right) \nn
&=& \rho  -\vp^2\frac{t_{\rm h}^2}{\pi U^2\rho^2}\int^\infty_0 d |\vk| |\vk|^3 (\nf(\ef)+\nb(\eb))\nn
&\approx& \rho  -\vp^2\frac{4\pi t_{\rm h}^2 \rhof^2}{ U^2\rho^2}.
\eeq
These two expressions (\ref{eq:alpha-expression}) and (\ref{eq:residue-expression}) determine the spectral properties of the Goldstino at small momentum. 
The last approximate equalities in the equations above are valid at small temperature, when $\int^\infty_0 d|\vk| |\vk|^3 \nb(\eb)$ can be neglected. 
The dispersion relation is quadratic~\cite{Shi:2009ak, Bradlyn:2015kca}, but the coefficient of  $\vp^2$,  $\alpha$, is different from that obtained from Eq.~(\ref{eq:result-dispersion-assumption}) under the assumption that the Goldstino pole exhausts the sum rule:
The last term in Eq.~(\ref{eq:alpha-expression}) does not exist in Eq.~(\ref{eq:result-dispersion-assumption}).
In fact, at weak coupling, the Goldstino ceases to exhaust the sum rule as soon as its momentum becomes finite.  
Equation  (\ref{eq:residue-expression}) reveals  indeed that the residue decreases rapidly with increasing $|\vp|^2$, from its maximum value  $Z=\rho$ achieved at $|\vp|=0$. The Goldstino continues to exist as a pole, and 
as can be seen in Fig.~\ref{fig:support-pole}, left panel, its quadratic dispersion being accurately reproduced with the value of $ \alpha$ given in Eq.~(\ref{eq:alpha-expression}), for  $|\vp|\lesssim 0.02$. 

\begin{figure*}[!t] 
\begin{center} 
\includegraphics[width=0.75\textwidth]{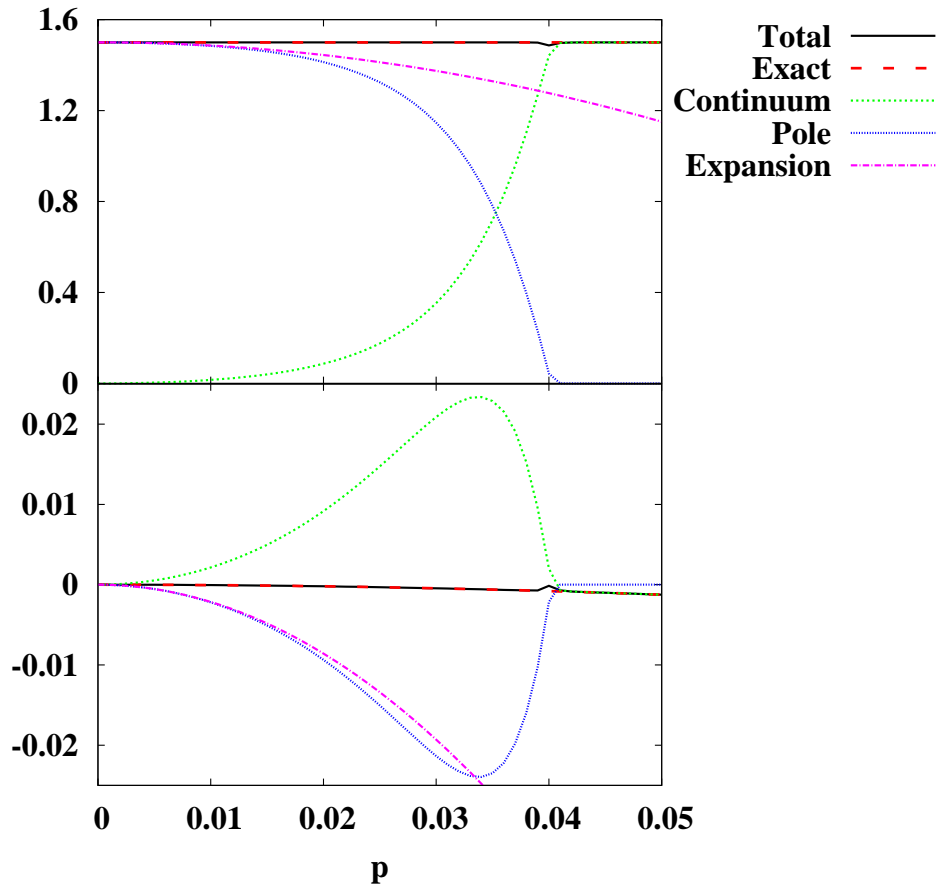}
\caption{{{(Color online)}}
 The contributions from the pole (``Pole'') and the continuum (``Continuum'') to the sum rules, Eqs.~(\ref{eq:sum-rule}) (upper panel) and (\ref{eq:sumrule-2nd-2}) (lower panel).
Their sum (``Total'') is also plotted.
For comparison, the pole contribution obtained from the small $|\vp|$ and $\tilde{\omega}$ expansion (``Expansion''; Eqs.~(\ref{eq:residue-expression}) and (\ref{eq:sumrule-2nd-pole})), and the right-hand sides (``Exact'') of these two sum rules are also plotted. Note that the curve ``Exact'' and ``Total'' coincide.
{{The unit of the vertical axis in the lower panel is $t_h$.
}}
} 
\label{fig:sumrules}
\end{center}
\end{figure*}

 The spectral strength missing in the residue   is moved to the continuum, in agreement with the sum rule, Eq.~(\ref{eq:sum-rule}). This is illustrated by the plot of the spectral function (the upper-left panel of Fig.~\ref{fig:spectrum-T=0}), while the respective contributions of the pole and the continuum to the sum rule are displayed in Fig.~\ref{fig:sumrules}, upper panel. This figure confirms that 
for $|\vp|$ smaller than 0.02, the spectral weight of the pole is well described by the result of small $|\vp|$, small $\tilde{\omega}$ expansion, Eq.~(\ref{eq:residue-expression}).
At such small $|\vp|$,  the spectral weight of the continuum is almost negligible, but it increases gradually as $|\vp|$ increases. 
Above $|\vp|=0.04$, where the pole hits the continuum, all the spectral weight is moved to the continuum. 

A similar analysis can be performed for the second sum-rule,  Eq.~(\ref{eq:sumrule-2nd}).
For convenience, we rewrite this sum rule in terms of $\tilde{\omega}$: 
\beq
\label{eq:sumrule-2nd-2}
\int_{-\infty}^{\infty}\frac{\rmd\tilde{\omega}}{2\pi} \,\tilde{\omega}\sigma(\tilde{\omega},\vp)
&=&- {t_{\rm h}}(\rhob-\rhof)\vp^2.
\eeq
The contribution from the pole,
\beq
\label{eq:sumrule-2nd-pole}
\int^\infty_{-\infty} \frac{\rmd \tilde{\omega}}{2\pi}   \,\tilde{\omega}\sigma(\tilde{\omega},\vp)
\simeq - t_{\rm h} \left\{(\rhob-\rhof)+ \frac{4\pi t_{\rm h}\rhof^2}{U\rho}\right\} \vp^2,
\eeq
 and that from the continuum are shown in the lower panel of Fig.~\ref{fig:sumrules}. 
 We note that the first two terms in the right hand side of Eq.~(\ref{eq:sumrule-2nd-pole}), namely $-t_{\rm h}(\rho_b-\rho_f)\vp^2$,  agree with the exact result, so we expect the last term  to cancel the continuum contribution.
 We also see that, at weak coupling, the latter term is much larger than the former ones.
 The figure illustrates that such a large cancellation indeed takes place between the contribution of the pole and that of the continuum, which can be explained by using the following estimates:  
The pole has a large strength ($\rho\sim{\cal{O}}(\vp^0)$), but its energy $\tilde\omega$ is small ($-\alpha\vp^2\sim{\cal{O}}(\vp^2)$) at small $|\vp|$.
The continuum on the other hand, has a small strength,  $-\vp^2 4\pi t_{\rm h}^2 \rhof^2/(U^2\rho^2)\sim{\cal{O}}(\vp^2)$, but a large energy ($U\rho\sim{\cal{O}}(\vp^0)$). 
The net result is a nearly complete cancellation  between pole and continuum contributions, leaving a sum rule that almost does not change with momentum (right-hand side of Eq.~(\ref{eq:sumrule-2nd-2})), while the location of the pole varies rapidly with $\vp$. In such circumstances, one cannot use the sum rule to determine the energy of the collective mode, as suggested by Eq.~(\ref{eq:result-dispersion-assumption}).\\

Above the momentum where the pole hits the continuum, the spectral function exhibits a very broad peak (see the upper-right panel of Fig.~\ref{fig:spectrum-T=0})).
As the momentum increases, the peak sharpens, as can be seen in  the lower-left panel of Fig.~\ref{fig:spectrum-T=0}).
Eventually, its position becomes almost the same as in the non  interacting limit, i.e., $\tilde\omega\simeq -t_{\rm h} \vp^2$ (Eq.~(\ref{eq:dispersion-largep}) with $\Delta\mu^0$ replaced by $\Delta\mu$). This is expected: 
In the limit where $\tilde{\omega}$ and $|\vp|^2$ are large compared with $U\rho$, $G^R(\omega,\vp)$ approaches Eq.~(\ref{eq:goldstino-one-loop}). When the momentum exceeds some value close to  the Fermi momentum, the pole emerges from the continuum, as is shown in the right panel of Fig.~\ref{fig:support-pole} (for the parameters used, the two curves in the right panel of Fig.~\ref{fig:support-pole}, corresponding respectively  to the actual pole position and to $\tilde{\omega}=-t_{\rm h}\vp^2$, are almost indistinguishable.) 
The excitation then corresponds to $\omega<0$, and  is therefore  induced by the operator $q_\vp$ that creates  fermion particle-boson  hole excitations. Before that point the   dispersion relation corresponds to a boson particle - fermion hole excitation, that is, an excitation induced by $q^\dagger_\vp$.

In summary, we have  described  the behavior of the Goldstino excitation from low to high momentum, in a generic weak coupling situation. As very small momentum the Goldstino exists as a well isolated excitation, with a single peak dominating the spectral function. As the momentum increases, the spectral weight of the peak decreases and is moved to the continuum of fermion hole excitations. At some value of the momentum, the peak merges with the continuum, producing a broad structure in the spectral function. This structure eventually turns again  into a sharp peak as the nature of the excitation gradually changes. The change is complete when the momentum is sufficiently large for the peak to emerge from the continuum, and $\omega$ becomes negative, at which point the excitation is essentially a boson hole excitation.  Note that both at small momentum and at high momentum, the excitation has a well defined quadratic dispersion relation, $\omega=\Delta\mu-\alpha \vp^2$ at small momentum, and $ \omega\approx -t_{\rm h} \vp^2$ at high momentum. \\

\begin{figure*}[!t] 
\begin{center} 
\includegraphics[width=0.75\textwidth]{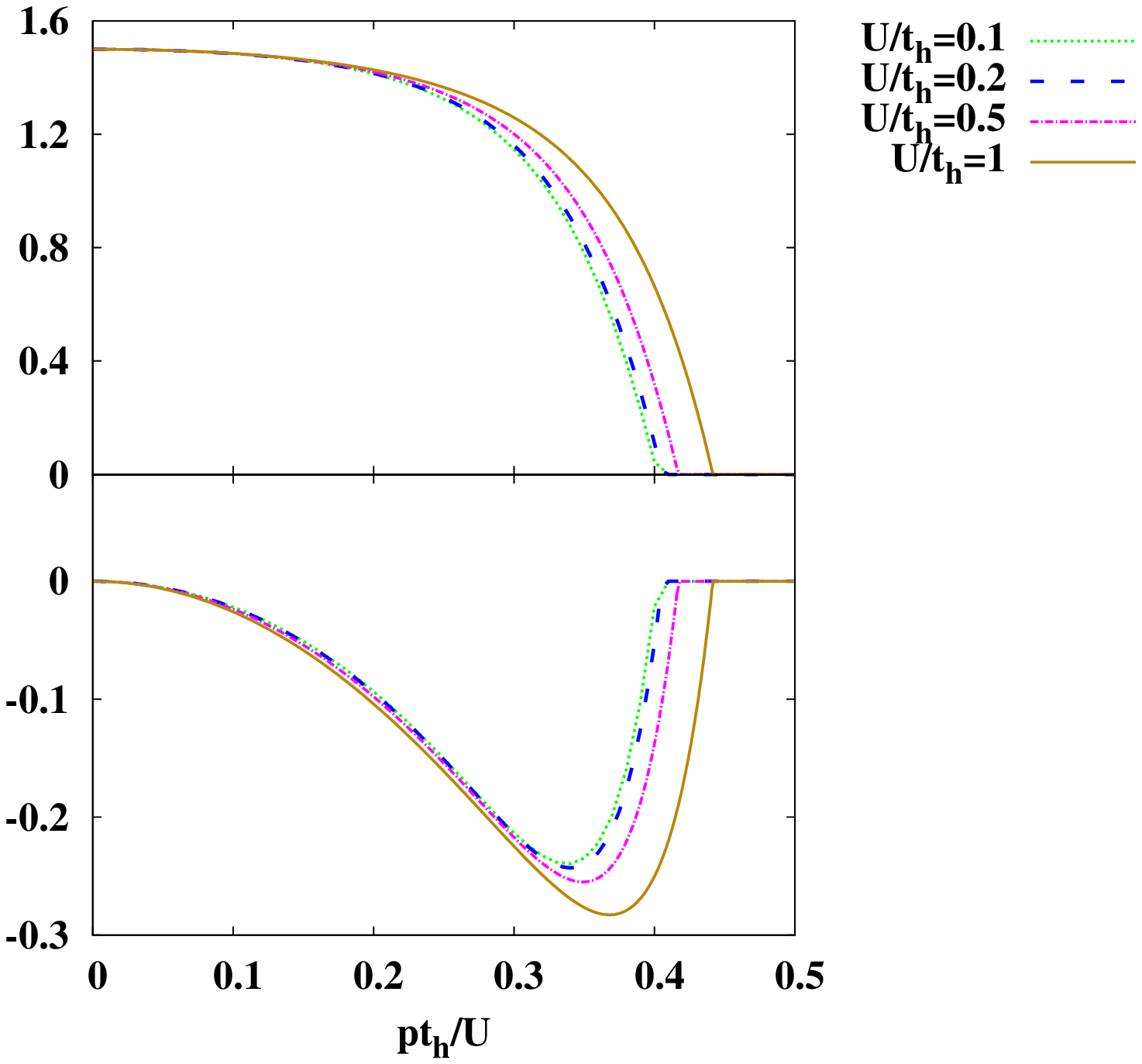}
\caption{{{(Color online)}}
 The contributions from the pole to the sum rules, Eqs.~(\ref{eq:sum-rule}) (upper panel) and (\ref{eq:sumrule-2nd-2}) (lower panel) at $U/t_{\rm h}=0.1$, $0.2$, $0.5$, and $1$.
In the lower panel, the vertical axis is also scaled by $U$ to show the invariance:
The quantity plotted in this panel is $\int \rmd\tilde{\omega} \,\tilde{\omega}\sigma(\tilde{\omega},\vp)t_{\rm h} /(2\pi U)$.
{{The unit of the vertical axis in the lower panel is $t_h$.
}}
}
\label{fig:sumrules-differentU}
\end{center}
\end{figure*}

Finite temperature produces a modest smearing, without altering the global structure of the results, at least  within the range that we have explored, i.e. $T/t_{\rm h} \le 1$. This can be seen on the plot of the  spectral function at $T/t_{\rm h}=0.2$ and $T/t_{\rm h}=1$ in Fig.~\ref{fig:spectrum-T=0}.
 Finite $T$ also increases the value of $\alpha$, as can be expected from Eq.~(\ref{eq:alpha-expression}):
 The integral appearing in the second line of this equation is dominated by large momenta, and increases when $T$ increases. 
The value of $\alpha$ increases by $\sim 20\%$ at $T=t_{\rm h}$ compared to its value at $T=0$.
 \\

Finally, let us see how the results are modified as one increases the coupling $U$. The left hand side of the sum rules, Eq.~(\ref{eq:sum-rule}) and Eq.~(\ref{eq:sumrule-2nd})  are plotted in Fig.~\ref{fig:sumrules-differentU} for the values  $U/t_{\rm h}=0.1$, $0.2$, $0.5$, and $1$. A comparison with Fig.~\ref{fig:sumrules}  reveals that, in this range of couplings,  the qualitative behaviors do not change much. 
In fact, because in the dominant terms of Eq.~(\ref{eq:residue-expression}),  $|\vp|$ appears only in the combination $|\vp|/U$, the sum rules do not change, when these are plotted as a function of $|\vp|t_{\rm h}/U$, as shown in  Fig.~\ref{fig:sumrules-differentU}.

\begin{figure*}[!t] 
\begin{center} 
\includegraphics[width=1.0\textwidth]{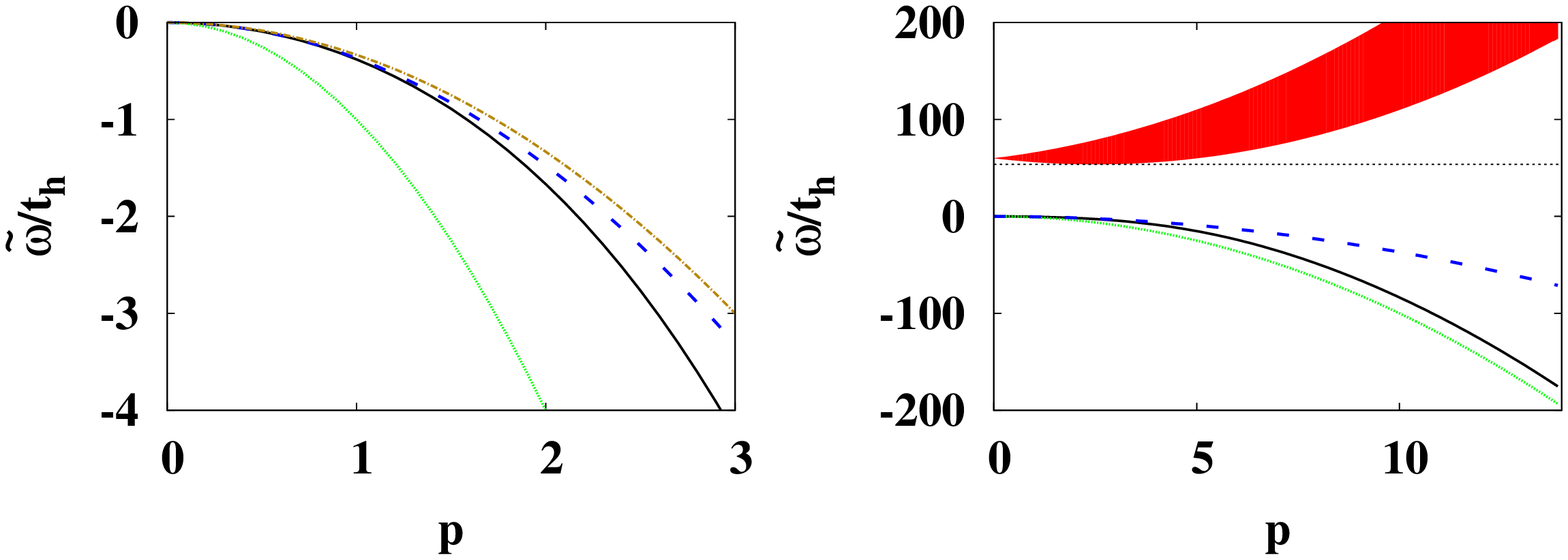}
\caption{ {{(Color online)}}
 The range of the continuum (shaded area) and the position of the goldstino pole (solid and black line) at small $|\vp|$ (left panel) and large $|\vp|$ (right panel) at $U/t_{\rm h}=40$.
For comparison, the asymptotic result on the pole position $\tilde{\omega}=-\alpha|\vp|^2$ at small $|\vp|$ is also plotted with the dashed (blue) line, the result $\tilde{\omega}=-t_{\rm h}|\vp|^2$ at large $|\vp|$ is plotted with the dotted (green) line, and $\tilde{\omega}=-\alpha_s\vp^2$ is plotted with the chain (dark yellow) line. 
The dashed horizontal line in the right panel indicates the value $\tilde\omega=-\Delta\mu$, corresponding to $\omega=0$. 
} 
\label{fig:support-pole-largeU}
\end{center}
\end{figure*}

\begin{figure*}[!t] 
\begin{center} 
\includegraphics[width=0.6\textwidth]{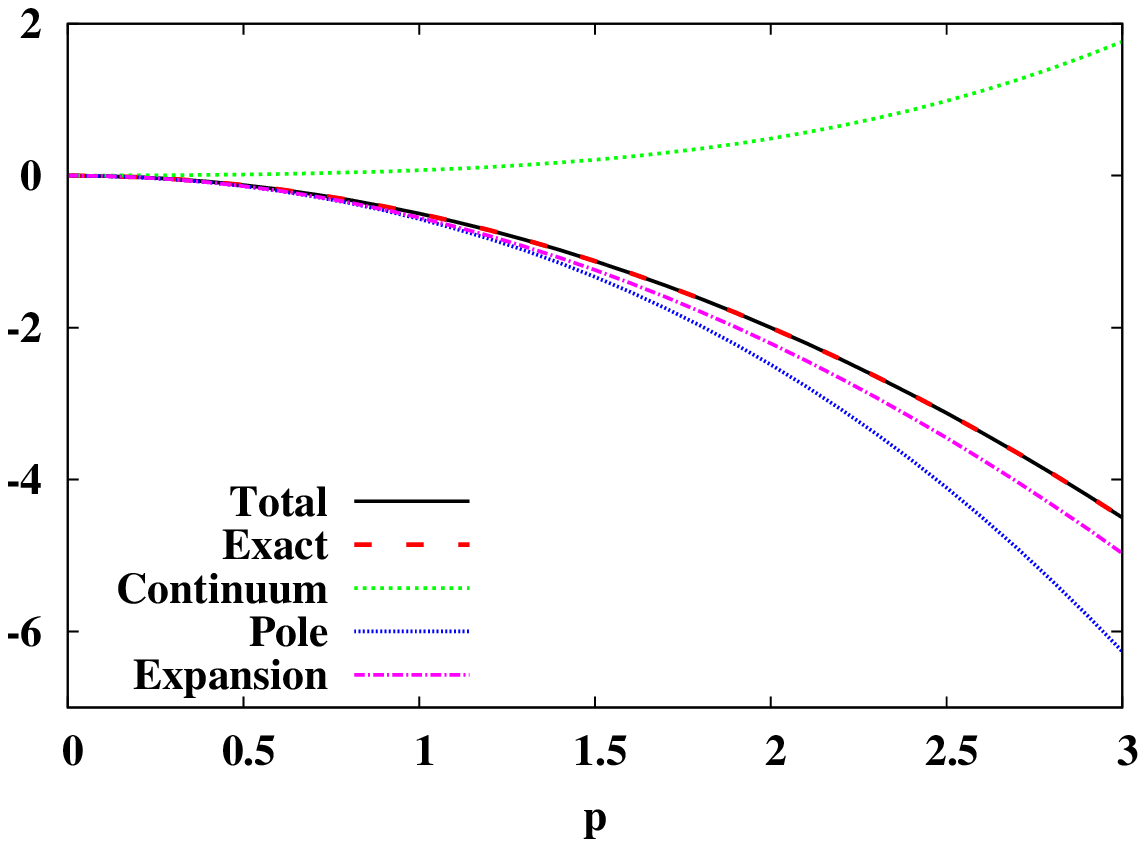}
\caption{{{(Color online)}}
The pole and continuum contributions to the sum rule (\ref{eq:sumrule-2nd-2}) for $U/t_{\rm h}=40$.
Their sum, the pole contribution obtained from the small $|\vp|$ and small $\tilde{\omega}$ expansion,  and the right-hand side of Eq.~(\ref{eq:sumrule-2nd-2}), labelled ``Exact'' are also plotted for comparison.
{{The unit of the vertical axis is $t_h$.
}}
} 
\label{fig:sumrule-2nd-largeU}
\end{center}
\end{figure*}

At much stronger coupling ($U/t_{\rm h}\gg 1$), the situation changes however, as illustrated in the left panel of Fig.~\ref{fig:support-pole-largeU}. 
In this case, the shift of the continuum $U\rho$ becomes large enough for the pole to stay away from the continuum. 
Nevertheless we still observe the change of the dispersion relation from $\tilde{\omega}=-\alpha \vp^2$ (small $|\vp|$) to $\tilde{\omega}=-t_{\rm h} \vp^2$ (large $|\vp|$). As can be seen in Fig.~\ref{fig:support-pole-largeU}, the dispersion relation is approximately $\tilde{\omega}=-\alpha \vp^2$  below $|\vp|\simeq 2$, and  approaches $\tilde{\omega}=-t_{\rm h} \vp^2$  above $|\vp|\simeq 12$.
At the same time, the residue  is almost equal to $\rho$ below $|\vp|\simeq 2$,  while it is approximately $\rho_b$ above $|\vp|\simeq 12$.
We note that the large value of $U$ makes $\Delta\mu$ negative, so the excitations that we are considering are those  induced by $q_\vk$: these are the excitations expected in the absence of the Fermi sea, and indeed at strong coupling, the Fermi sea provides just a renormalization of the dispersion relation, without affecting much the structure of the excitation.
Furthermore, when $U$ is large, the  $U$-independent contribution to  $\alpha$ in Eq.~(\ref{eq:alpha-expression}) dominates over the $U$-dependent term, so that $\alpha$ approaches $\alpha_{\rm s}$ defined by Eq.~(\ref{eq:alpha-expression-largeU}), as can be seen from Fig.~\ref{fig:support-pole-largeU}.
This means that in this regime (large coupling and small momentum), the dispersion relation is completely determined by the sum rule of Eq.~(\ref{eq:result-dispersion-assumption}). Actually, one can gauge the validity of this assumption by looking at Fig.~\ref{fig:sumrule-2nd-largeU} where the various contributions to the sum rule are displayed:
In contrast to Fig.~\ref{fig:sumrules}, the cancellation between the pole and the continuum is quite modest, and the contribution from the pole is almost equal to the right-hand side of Eq.~(\ref{eq:sumrule-2nd-2}).

\section{Conclusions and outlook}
\label{sec:summary}

We have analyzed the spectral properties of the Goldstino that can be produced in a 2-dimensional system of cold atoms and molecules with supersymmetric dynamics. 
The supersymmetry is spontaneously broken by the finite density of particles, and it is also explicitly broken by the difference $\Delta\mu$ of the chemical potentials of fermions and bosons: this is why, at zero momentum, the energy of the Goldstino is not zero, as would be expected from the Goldstone theorem, but equal to $\Delta\mu$. 
As soon as the momentum is finite, the fermions produce a significant broadening of the excitation: 
this is because, in contrast to what happens in usual collective modes, the fermion excitations are not concentrated in the vicinity of the Fermi surface, but the entire Fermi sea participates in the excitation. 
The situation where there are no fermion in the ground state is simpler, since then the Goldstino always appear as a sharp peak, and its properties can be simply obtained from the sum rules that we have derived. 
A similar situation holds at strong coupling. 
At weak coupling, the dispersion relation of the mode is quadratic when it is well defined, i.e. at small momentum ($\tilde\omega=-\alpha\vp^2$) and at large momentum  ($\tilde\omega=-t_{\rm h}\vp^2$). 
A smooth crossover from the small momentum region to the large momentum region was observed in analyzing the spectral function.
We note that this Goldstino corresponds  to the type-B mode in the classification of Nambu-Goldstone modes~\cite{Watanabe:2012hr, Hidaka:2012ym}
because the order parameter $\rho$ is expressed as the expectation value of the anti-commutation relation between the supercharge and its density, $\rho= \langle\{Q,q^\dag(\vx)\}\rangle$. The quadratic dispersion is consistent with the classification for the internal symmetry breaking.

There are a number of interesting issues which would be worth pursuing. The transition between weak and strong coupling that we just alluded to is one of them. It would also be interesting to see how the results obtained in this paper are modified by the presence of a confining trap. More generally, it has been shown  that  the dispersion relation of (bosonic) Nambu-Goldstone modes can be determined in a model-independent way~\cite{Hidaka:2012ym, Watanabe:2012hr}.  It would be interesting to investigate whether the techniques of these papers can be applied to the Goldstino. Studying the effect of other explicit forms of symmetry breaking (beyond that due to the chemical potential difference) would be worth exploring.
The damping rate of the Goldstino cannot calculated in the approximation used in the present paper (beyond the trivial broadening that we have discussed),  but improved approximations that take into account the collisions  would enable us to calculate this quantity  \cite{Lai:2015fia}.
Another topic of interest is the study of the Goldstino in cases where the bosons form a Bose-Einstein condensate.
We hope to return to these and other interesting questions in the near future. 

\section*{Acknowledgement}

This work was supported by the Grant-in-Aid for the Global COE Program ``The Next Generation of Physics, Spun from Universality and Emergence'' from the Ministry of Education, Culture, Sports, Science and Technology (MEXT) of Japan.
This work was also supported by JSPS KAKENHI Grants Number 15H03652, the RIKEN iTHES Project, and JSPS Strategic Young Researcher Overseas Visits Program for Accelerating Brain Circulation (No. R2411).
D. S. thanks T. Kunihiro for encouraging this research, and I. Danshita and Y. Takahashi for fruitful discussion. JPB's research is 
supported by the European Research Council under the
Advanced Investigator Grant ERC-AD-267258. JPB thanks the Yukawa Institute of Theoretical Physics in Kyoto for hospitality in the summer 2015, during which this paper was completed.\\


\end{document}